\documentclass[epj]{svjour}
\usepackage{graphics}
\usepackage{bm}       % bold math
\newcommand{\hfbax}{\sc hfb-ax}
\newcommand{\hfbtho}{\sc hfbtho}
\newcommand{\rr} {\boldsymbol{r}}
\newcommand{\bea}{\begin{eqnarray}}
\newcommand{\eea}{\end{eqnarray}}
\begin{document}
\title{Coordinate-Space Hartree-Fock-Bogoliubov Description of
Superfluid Fermi Systems}
\author{J.C. Pei\inst{1-3}, W. Nazarewicz\inst{2-4} \and
 M. Stoitsov\inst{2,3,5}
% \thanks is optional - remove next line if not needed
%\thanks{\emph{Present address:} Insert the address here if needed}%
}                     % Do not remove
%
%\offprints{}          % Insert a name or remove this line
%
\institute{Joint Institute for Heavy Ion Research, Oak Ridge
National Laboratory, Oak Ridge, TN 37831, USA
 \and
 Department of Physics and
Astronomy, University of Tennessee Knoxville, TN 37996, USA
\and
Physics Division, Oak Ridge National Laboratory, P.O. Box 2008, Oak
Ridge, TN 37831, USA
\and
Institute of Theoretical Physics, Warsaw University, ul.Ho\.{z}a 69,
PL-00681 Warsaw, Poland
\and Institute of Nuclear Research and Nuclear
Energy, Bulgarian Academy of Sciences, Sofia, Bulgaria
}
\date{Received: date / Revised version: date}
% The correct dates will be entered by Springer
%
\abstract{Properties of strongly interacting, two-component
finite Fermi systems are
discussed within the   recently developed coordinate-space
Hartree-Fock-Bogoliubov (HFB) code {\hfbax}. Two illustrative
examples are presented:
(i) weakly bound deformed Mg isotopes, and (ii) spin-polarized
 atomic condensates in a  strongly deformed harmonic trap.
\PACS{
      {21.60.Jz}{Nuclear Density Functional Theory}   \and
      {31.15.Es}{Atomic Density Functional Theory }
     } % end of PACS codes
} %end of abstract
\maketitle
\section{Introduction}
\label{intro} Superconductivity and Cooper pair formation   are
generic features of strongly-interacting
 many-body Fermi systems.  In the context of the
Density Functional Theory (DFT), the Hartree-Fock-Bogoliubov (HFB or
Bogoliubov-de-Gen\-nes) framework has been widely used to treat
pairing correlations in nuclei
(see, e.g., \cite{dobaczewski96,stoitsov05,Bertsch}) and ultracold atom
gases (see, e.g., \cite{Bertsch,sensarma,kinnunen,bulgac08,Bulgac07}). The superiority
of the HFB method
over  the conventional BCS approximation becomes particularly  apparent
in the context of weakly bound systems, such as drip-line nuclei, where
the coupling to the scattering continuum becomes essential~\cite{dobaczewski96}.

The HFB equations can be solved in several ways (see Ref.~\cite{jcpei}
for a recent overview). In the commonly used
configuration
space approach, the quasi-particle orbitals are expanded in a suitable
single-particle basis.  A number of  HFB codes were  developed
by employing  the harmonic oscillator (HO) eigenstates
\cite{stoitsov05,dobaczewski04}.
However,  use of the HO
basis is questionable in the limit of both weak binding and very
large deformations, which  require the use of unrealistically large
configuration spaces to guarantee convergence. In both situations,
the coordinate-space approach to the HFB problem
\cite{dobaczewski96,bulgac80,dobaczewski84} is superior.

A number of coordinate-space techniques have been
developed over the years, and their performance strongly depends on
the size and self-consistent symmetries of the spatial mesh employed~\cite{blum,terasaki}.
While the direct iterative diagonalization procedure in the coordinate space is
computationally intensive,
the  advent  of teraflop
supercomputers enables us to carry out  large-scale DFT
computations of complex physical systems in non-spherical spatial boxes. The
recently developed parallel 2D-HFB solvers utilizing  the B-spline technique
offer excellent accuracy when
describing  deformed weakly bound nuclei~\cite{teran,jcpei}.
Solving the HFB equations in a 3D coordinate space is not a simple
task; it is worth noting that several developments
are underway, such as a general-purpose 3D-HFB solver based on
multi-resolution analysis and wavelet expansion~\cite{fann,jcpei}.

We recently released a 2D coordinate-space  code {\hfbax} that solves
the HFB problem  using B-splines~\cite{jcpei}. Its high precision has been
explicitly demonstrated by
testing against the HO basis expansion method, wavelet method,  and other
HFB codes.
In this work, we apply  {\hfbax} to two problems of
current interest. First, we study the drip-line Mg isotopes using the
SLy4~\cite{sly4} energy density functional. The heaviest-known  $^{40}$Mg isotope
has recently been produced~\cite{Baumann}, and is expected
to be  weakly deformed~\cite{stoitsov03}. Here,
we systematically compare the differences between HO
expansion and coordinate-space HFB calculations.

In addition to nuclear calculations, {\hfbax}  has recently
been applied to
cold atomic systems within the Superfluid Local Density
Approximation (SLDA)~\cite{Bulgac07,bulgac}. Studies of
pairing properties of polarized cold  atom gases are indeed of considerable
experimental~\cite{partridge,ketterle} and theoretical \cite{sensarma,kinnunen,bulgac08}
interest.  The
separation between paired and normal phases
has been observed in a  gas with unequal numbers of two
spin components trapped in an extremely deformed
potential~\cite{partridge}. Calculations indicate that experimental
results depend on the number of fermions and  trap
asymmetry~\cite{sensarma}. Theoretical simulations based on the HO basis
method are, however,  limited to spherical shapes or small deformations \cite{sensarma,kinnunen}.
In this work, we calculate the properties of a Fermi gas of 1000 polarized atoms
trapped in an extremely elongated  HO potential.

\begin{table*}[htb]
 \caption{\label{tabho1}
 Comparison between {\hfbax} and {\hfbtho} with SLy4 p-h
 functional and mixed  pairing for neutron-rich even-even nuclei $^{34-44}$Mg.  All energies are in
 MeV. The quadrupole moments $Q_{20}$ are in fm$^2$. See text
 for more details.}
\hspace{2cm}\begin{tabular}{ccccccccc} \hline\noalign{\smallskip}
Nuclei & {\hfbax} & {\hfbax} & {\hfbtho}& {\hfbax} & {\hfbtho} & {\hfbax}   & {\hfbtho} & {\hfbax}\\
& $Q_{20}$&$\lambda_n$&$E_{tot}$ & $E_{tot}$& $E_{kin}$&$E_{kin}$&$E_{pair}$&$E_{pair}$\\
 \hline
$^{34}$Mg &+148.0&$-$3.63&$-$257.49&$-$257.48&553.51&553.54&$-$5.28&$-$3.98\vspace{1pt}\\
$^{36}$Mg &+175.4&$-$2.49&$-$263.02&$-$263.06&585.89&585.63&$-$0.02&$-$0.12 \vspace{1pt}\\
$^{38}$Mg &+187.9&$-$1.37&$-$265.40&$-$265.46&626.39&625.16&$-$2.49&$-$2.59\vspace{1pt}\\
$^{40}$Mg &+221.6&$-$0.45&$-$266.96&$-$267.13&655.66&653.98& 0 &0 \vspace{1pt}\\
$^{42}$Mg &$-$136.1&$-$0.22&$-$264.83&$-$264.91&699.68&695.00&$-$5.11&$-$4.34\vspace{1pt}\\
$^{44}$Mg&$-$107.8&+0.15&$-$264.19&$-$264.56&730.35&723.02&$-$2.68&$-$3.44\\
\noalign{\smallskip}\hline
\end{tabular}
\end{table*}

\section{Nuclear calculations}
\label{sec:1}

In the coordinate-space representation,
the HFB equations
 can be
written as~\cite{dobaczewski96,stoitsov05}:
\begin{equation}\label{HFBm}
\begin{array}{c}
\displaystyle \int d\mathbf{r'} \displaystyle \sum_{\sigma'} \left(
\begin{array}{cc}
h(\mathbf{r}\sigma,\mathbf{r'}\sigma')-\lambda&\tilde{h}(\mathbf{r}\sigma,\mathbf{r'}\sigma') \vspace{3pt} \\
\tilde{h}(\mathbf{r}\sigma,\mathbf{r'}\sigma') &-h(\mathbf{r}\sigma,\mathbf{r'}\sigma')+\lambda\\
\end{array}
\right) \vspace{5pt} \\   ~~~~~~~~~~ \times\left(
\begin{array}{c}
\psi^{(1)}(\mathbf{r'}\sigma') \vspace{2pt}\\
\psi^{(2)}(\mathbf{r'}\sigma') \\
\end{array}
\right)=E\left(
\begin{array}{c}
\psi^{(1)}(\mathbf{r}\sigma) \vspace{2pt}\\
\psi^{(2)}(\mathbf{r}\sigma) \\
\end{array}
\right),
\end{array}
\end{equation}
where ($\mathbf{r}, \sigma$) are the particle spatial and spin
coordinates, $h(\mathbf{r}\sigma,\mathbf{r'}\sigma')$ and
$\tilde{h}(\mathbf{r}\sigma,\mathbf{r'}\sigma')$ are the
particle-hole (p-h) and particle-particle (p-p) components of the
HFB  Hamiltonian, respectively,
$\psi_n^{(1)}(\mathbf{r}\sigma)$ and
$\psi_n^{(2)}(\mathbf{r}\sigma)$ are the upper and lower components
of the  HFB wave function, and $\lambda$ is the
chemical potential. The spectrum of quasiparticle energies $E$ is
discrete for $|E|<-\lambda$ and continuous for $|E|>-\lambda$. By
requiring that the eigenfunctions vanish at the edge of the box (box
boundary conditions), the particle continuum becomes discretized.

In the axial  geometry, the third component of the
single-particle angular momentum, $\Omega$, is a good quantum
number. The HFB wave function can thus be written as
$\Psi_n^{\Omega q}({\rr})$ where ${\rr}=(\phi, \rho,
z)$, $q$=$\pm\frac{1}{2}$ denotes the cylindrical isospin
coordinates, and $\Omega$=$\pm\frac{1}{2}$, $\pm\frac{3}{2}$,
$\pm\frac{5}{2}$, $\ldots$. We also assume that the reflection symmetry
is conserved; hence
the parity $\pi$ is a good quantum number.
Consequently, the Hamiltonian matrix becomes block diagonal, which enables parallelization.

 In
nuclear calculations, the p-h channel is often modeled with the
Skyrme energy density functional, while a zero-range  isovector
pairing interaction is often used in the p-p channel. By using the
zero-range interactions, Eq.~(\ref{HFBm}) becomes local and
easier to solve. In this study, we used the  SLy4~\cite{sly4} Skyrme functional
 and the
mixed $\delta$ pairing \cite{dobaczewski02}. The pairing strength
has been  fitted to reproduce the average neutron pairing gap in
$^{120}$Sn~\cite{dobaczewski04}. [The numerical effort depends on
the box size ($\rho_{\rm max}$, $z_{\rm max}$), the largest distance
between neighboring mesh points in the grid $h$ (the B-spline grid
is not uniform), and the order of B-splines $M$]. We
demonstrated~\cite{jcpei} that   $h$=0.6\,fm and $M$=13 guarantee
excellent precision of calculations. For heavy elongated nuclei,
such as those on the way to fission,
 large 2D  boxes are required. To accelerate convergence of self-consistent
 iterations, we employed the modified
Broyden mixing~\cite{baran} which is significantly faster than the standard
linear mixing method. In the present calculations for the  Mg isotopes, we
used a box of $\rho_{max}$=$z_{max}$=18 fm.

Table~I displays the results of  calculations for $^{34-44}$Mg
isotopes  with  {\hfbtho} \cite{stoitsov05} and {\hfbax}. Our {\hfbtho} calculations
were carried out in a configuration space of 20 HO shells. We
compare the total binding energy $E_{tot}$, kinetic energy $E_{kin}$,
 pairing energy $E_{pair}$,
 total  quadrupole moment $Q_{20}$, and neutron chemical potential
$\lambda_n$. We can see that while the two codes yield
 very similar total binding energies, their differences increase
towards the neutron drip line. For example, in
$^{34}$Mg the difference in $E_{tot}$ is only 10\,keV between the two codes,
while it becomes 170\,keV in
  $^{40}$Mg, As discussed in \cite{jcpei},
the kinetic and pairing energies are slightly different in
{\hfbtho} and {\hfbax} because of different continuum discretization:
 {\hfbtho} predicts  kinetic
energies that are systematically larger than in {\hfbax}, especially for nuclei near drip lines.

In our calculations with SLy4, $^{40}$Mg is the last  Mg isotope that is  stable
against the two-neutron emission. This is consistent with
recent experimental evidence~\cite{Baumann}. According to  Table~I,
its ground state is predicted  to have a well-deformed, prolate shape.
(A deformed halo structure in $^{40}$Mg has also been predicted in
Ref.~\cite{nakada}.) Interestingly, due to deformed shell gaps,
static proton and neutron pairing  vanishes in this nucleus.
The heavier even-even
isotope of
$^{42}$Mg is predicted to have a negative neutron chemical potential
 but is unbound to two-neutron emission.
 As this nucleus is expected to have appreciable oblate deformation, however,
the particle stability
of $^{42}$Mg could be enhanced against the two-neutron
decay~\cite{stoitsov03}.

\section{Atomic calculations}

The strongly interacting,  polarized Fermi gas can be described by
introducing  mismatched chemical potentials,
$\lambda_{\uparrow}$ and $\lambda_{\downarrow}$,  corresponding to spin-up (majority)
and
spin-down (minority)  states, respectively. The  HFB equations of the resulting
Two-Fermi Level Approach (2FLA)
 read \cite{sensarma}:
 \bea
  \left(
\begin{array}{clrr}%
 H_0(\rr)-\lambda_{\uparrow} & {\hspace{0.7cm} } \Delta(\rr) \\
 \Delta^*(\rr) & -H_0(\rr) +\lambda_{\downarrow}
\end{array}
\right)\left(
\begin{array}{clrr}%
u_i(\rr) \\
v_i(\rr)
\end{array}
\right)=E_i\left(
\begin{array}{clrr}%
u_i(\rr) \\
v_i(\rr)
\end{array}
\right)\label{BdG} \eea
As usual, we define the average chemical potential
$\lambda=(\lambda_{\uparrow}+\lambda_{\downarrow})/2$ and the difference
$2\lambda_s=\lambda_{\uparrow}-\lambda_{\downarrow}$. The diagonalization of
Eq.~(\ref{BdG}) is equivalent to Eq.~(\ref{HFBm}) by replacing the two chemical
potentials with $\lambda$, and adding a cranking term involving  spin projection.
As discussed in Ref.~\cite{Bertsch}, the 2FLA
is equivalent to the standard blocking procedure, and $\lambda_s$ can be viewed as
a rotational frequency that generates spin polarization.

The associated polarization density $m(\rr)=\rho_{\uparrow}(\rr)-\rho_{\downarrow}(\rr)$, total density $\rho(\rr)=\rho_{\uparrow}(\rr)+\rho_{\downarrow}(\rr)$,
pair density $\kappa(\rr)$, and
pairing potential $\Delta(\rr)$ are:
\begin{eqnarray}
m(\rr)&=& \sum_{0 \le E_i < \lambda_s} \left ( |u_i(\rr)|^2 +   |v_i(\rr)|^2 \right ),  \label{mr}\\
\rho(\rr)&=&m(\rr) + \sum_{E_i>\lambda_s}2 |v_i(\rr)|^2, \label{rhor}\\
\kappa(\rr)&=&\sum_{E_i>\lambda_s} u_i(\rr)v_i^*(\rr), \label{kappar} \\
\Delta(\rr)&=&-g_{{\it eff}}(\rr)\kappa(\rr),
\end{eqnarray}
where $g_{eff}(\rr)$ is the regularized pairing
strength. The regularization procedure is introduced because the
kinetic and pairing energies diverge with energy for  zero-range
pairing forces \cite{bulgac,borycki}.

\begin{figure}[htb]
% Use the relevant command for your figure-insertion program
% to insert the figure file.
% For example, with the option graphics use
\hspace{10pt}
\resizebox{0.45\textwidth}{!}{%
  \includegraphics{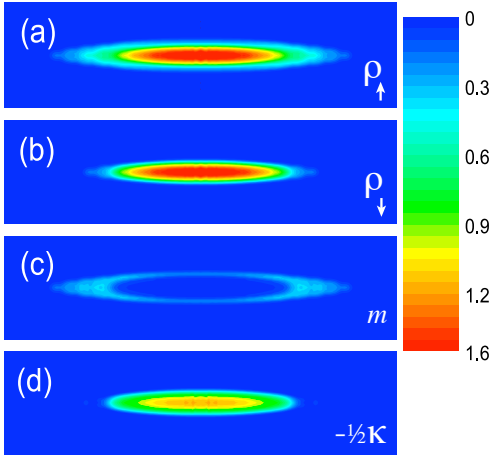}
}\hspace{-10pt}
\vspace{0.3cm}
\caption{(Color online) The calculated density contours of  Fermi
gas with a spin polarization of $P$=0.21: majority spin state density $\rho_{\uparrow}(\rr)$ (a);
minority spin state density $\rho_{\downarrow}(\rr)$ (b);  polarization density (c); and the pair density
$\kappa(\rr)$ (d).
The panels are plotted with
dimensionless length of 80 and width of 20. The corresponding density values
are given in the legend. To facilitate comparison, $\kappa$ is multiplied by a factor $-0.5$.
}
\label{fig:1}       % Give a unique label
\end{figure}
The energy density functional of SLDA  (in units $\hbar$=$m$=1) is given by
\cite{Bulgac07}:
\begin{equation}
{\cal E}(\rr)= \alpha \frac{\tau(\rr)}{2}
                 + \beta \frac{3(3\pi^2)^{2/3}\rho^{5/3} (\rr)}{10}
                 + g_{\it eff} \frac{|\kappa(\rr)|^2}{\rho^{1/3} (\rr)},
\end{equation}
where $\alpha$ and  $\beta$ are dimensionless parameters.
In the present calculation, we took the axial external HO trap
$V(\rho,z)=\frac{1}{2}m\omega^2(\rho^2+z^2/\eta^2)$. The parameter
$\eta$
defines the trap's an\-iso\-tropy and we took $\omega$=1.
In the experiment of Ref.~\cite{partridge}, an extremely large deformation
 of $\eta\simeq$50 was employed. Previous
theoretical work  based on HO expansion considered either a
spherical geometry \cite{kinnunen} or
a fairly small value of
$\eta\leq$4 \cite{sensarma}.
By taking advantage of the coordinate space formalism,
we could reach very large deformations of
$\eta$=10. Unlike in nuclei, the
spin-orbit coupling is absent in the atomic  problem. Therefore, the dimension of the
Hamiltonian matrix for the gas is  half   of the nuclear one, and  very
large spaces can be reached. We adopted a  rectangular box with
dimensionless $\rho_{max}$=10 and $z_{max}$=50 for the calculation
of 1000 atoms ($N$=$N_\uparrow$+$N_\downarrow$=1000). The
polarization, $P$=($N_\uparrow-N_\downarrow$)/$N$, is realized by
adjusting the value of $\lambda_s$. The resulting quasi-particle atomic
spectrum is much denser than a typical nuclear one.
A finer mesh size $h$=0.4 than in nuclear calculations,
and $M$=13 order B-splines guarantee convergence of  our SLDA calculations.

\begin{figure}[htb]
% Use the relevant command for your figure-insertion program
% to insert the figure file.
% For example, with the option graphics use
\hspace{20pt}
\resizebox{0.4\textwidth}{!}{%
  \includegraphics{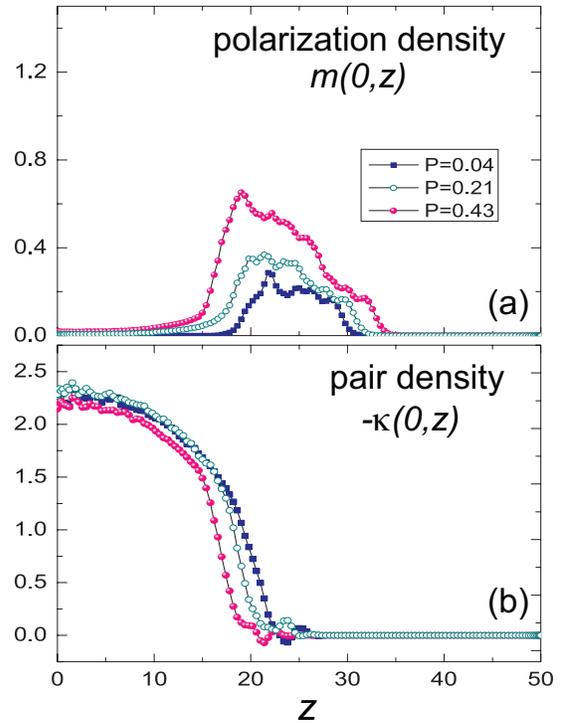}
}\hspace{-20pt}
\vspace{0.3cm}
\caption{(Color online) The polarization (a) and pair (b) densities of trapped  fermions
at different polarizations $P$.
The main effect of the polarization is  to decrease the superfluid core size.
}
\label{fig:2}       % Give a unique label
\end{figure}

Figure~\ref{fig:1}  shows the density contours for the case of 214 unpaired
particles ($P$=0.21).
The distribution of pair density $\kappa(\rr)$ coincides with that of
$\rho_{\downarrow}(\rr)$. That is,
 the spin-polarized  (unpaired) fermions shown in
Fig.~\ref{fig:1}(c)
 are distributed
around the paired superfluid core. This spatial separation between
normal and superfluid phases
is similar to what has been seen experimentally \cite{partridge}.

The  densities $m(\rho=0,z)$ and $\kappa(\rho=0,z)$ corresponding to different polarizations
$P$ are displayed in Fig.~\ref{fig:2}. The effect of phase separation is clearly seen. As discussed in Ref.~\cite{kinnunen}, the main effect of polarization is
to decrease the size of the superfluid core. The superfluid-to-normal transition
is not sharp, as expected in the finite system. In the intermediate
 region of small (but nonzero) $\kappa$ and nonzero $m$, the pair density
exhibits small-amplitude oscillations. As discussed in Refs.~\cite{sensarma,kinnunen}, such behavior is characteristic
of Fulde-Ferrel-Larkin-Ovchinnikov (FFLO) phase \cite{fflo} associated with the {\em magnetized superfluid}.

\section{Summary}

We used the recently developed coordinate-space HFB code {\hfbax} to
study the strongly interacting superfluid Fermi systems, including
nuclei and cold atomic gases. We first calculated the properties of
weakly bound deformed Mg isotopes by solving Skyrme HFB equations. We
conclude that the two-neutron drip line in the Mg chain predicted with
SLy4 density functional corresponds to $^{40}$Mg, but $^{42}$Mg can be
long-lived due to shape coexistence effects. Secondly, we investigated
the density distributions of a polarized cold atomic gas in an extremely
deformed external trap. The calculated density profiles clearly show the
separation between the  paired and  polarized normal phases, in agreement with
experimental findings. In the intermediate magnetized superfluid region,
the pairing field exhibits small amplitude oscillations, consistent with
FFLO behavior. We conclude that the coordinate-space HFB framework  is a
very useful tool for the description of weakly bound and/or extremely
deformed and polarized superfluid Fermi systems.
\\

This work was supported in part by the U.S.~Department of Energy
under Contract Nos.~DE-FG02-96ER40963 (University of Tennessee),
DE-AC05-00OR22725 with UT-Battelle, LLC (Oak Ridge National
Laboratory), DE-FG05-87ER40361 (Joint Institute for Heavy Ion
Research), and DE-FC02-07ER41457 with UNEDF SciDAC Collaboration.
Computational resources were provided by the National Center for
Computational Sciences at Oak Ridge and the National Energy Research
Scientific Computing Facility.

% For one-column wide figures use
% BibTeX users please use
% \bibliographystyle{}
% \bibliography{}
%
% Non-BibTeX users please use

\end{document}